\documentclass[a4paper,12pt]{article}
\usepackage{amssymb,amsthm,latexsym}

\newtheorem{theorem}{Theorem}

\title{MATHEMATICAL ASPECTS OF MEAN FIELD SPIN GLASS THEORY}

\author{
Francesco Guerra\footnote{\
e-mail: {\tt francesco.guerra@roma1.infn.it}} \\
{\small {\itshape Dipartimento di Fisica, Universit\`a di Roma ``La Sapienza''}}
\\
{\small {\itshape and INFN, Sezione di Roma1, Piazzale A. Moro 2, 00185 Roma, 
Italy}}
} 
\date{\today}
\begin{document}
\maketitle

\begin{abstract}
A comprehensive review will be given about the rich mathematical structure
of mean field spin glass theory, mostly developed, until now, in the frame
of the methods of theoretical physics, based on deep physical intuition and 
hints coming from numerical simulation. 
Central to our treatment
is a very simple and yet powerful interpolation method, 
allowing to compare different
probabilistic schemes, by using convexity and positivity arguments. 
In this way we can prove the existence of the
thermodynamic limit for the free energy density of the system, 
a long standing open problem. Moreover, in the frame of a generalized 
variational principle, we can show the emergency
of the Derrida-Ruelle random probability cascades, leading to the form of 
free energy given by the celebrated Parisi \textit {Ansatz}. All these results seem to
be in full agreement with the mechanism of spontaneous replica symmetry
breaking as developed by Giorgio Parisi.
\end{abstract}
\newpage

\section{Introduction}

The mean field model for spin glasses, introduced by David Sherrington and 
Scott Kirkpatrick more that thirty years ago \cite{SK}, \cite{KS}, is a 
celebrated model. 
Hundreds and hundreds of articles have been devoted to its study during the 
years, appearing in the theoretical physics literature.

The relevance of the model stems surely from the fact that it is intended 
to represent some important features of the physical spin glass systems, 
of great interest for their peculiar properties, exhibiting a new magnetic 
phase, where magnetic moments are frozen into disordered equilibrium 
orientations, without any long-range order. See for example \cite{stein} 
for a very readable review about the physical properties of spin glasses.

But another important source of interest is connected with the fact that 
disordered systems, of the Sherrington-Kirkpatrick type, and their 
generalizations, seems to play a very important role for theoretical and 
practical assessments about hard optimization problems, as it is shown for 
example by Mark M\'ezard, Giorgio Parisi and Riccardo Zecchina in \cite{MPZ}.

It is interesting to remark that the original paper was entitled 
``Solvable Model of a Spin-Glass'', while a previous draft, as I have it 
from David Sherrington, contained even the stronger denomination ``Exactly 
Solvable''. However, it turned out that the very natural solution devised 
by the authors was valid only at high temperatures, or large external 
magnetic fields. While, at low temperatures, the proposed solution 
exhibited a nonphysical drawback given by a negative entropy, as properly 
recognized by the authors in their very first paper.

It took some years to find an acceptable solution. This was done by Giorgio 
Parisi in a series of papers, by marking a radical departure from the 
previous methods. In fact, a very deep method of ``spontaneous replica 
symmetry breaking'' was developed. As a consequence the physical content of 
the theory was encoded in a functional order parameter of new type, and a 
remarkable structure emerged for the pure states of the theory, a kind of 
hierarchical, ultrametric organization. These very interesting 
developments, due to Giorgio Parisi, and his coworkers, are explained in a 
lucid way in the classical book \cite{MPV}. Part of this structure will be 
recalled in the following.

It is important to remark that Parisi solution is presented in the form of 
an ingenious and clever \textit{Ansatz}. Until few years ago it was not known 
whether this \textit{Ansatz} would give the true solution for the model, in 
the so called thermodynamic limit, when the size of the system becomes 
infinite, or it would be only a very good approximation for the true 
solution.

The general structure offered by the Parisi solution, and their possible 
generalizations for similar models, exhibit an extremely rich and 
interesting mathematical content. Very appropriately, Michel Talagrand has 
inserted a strongly suggestive sentence in the title to his recent book 
\cite{T}:``Spin glasses: a challenge for mathematicians''.

As a matter of fact, how to face this challenge is a very difficult 
problem. Here we would like to recall the main features of a very 
powerful method, yet extremely simple in its very essence, based on a 
comparison and interpolation argument on sets of Gaussian random variables.

The method found its first simple application in \cite{Gsum}, where it was 
shown that the Sherrington-Kirkpatrick replica symmetric approximate 
solution was a rigourous lower bound for the quenched free energy of the 
system, uniformly in the size. Then, it was possible to reach a long 
waited result \cite{GTthermo}: the convergence of the free energy density in the 
thermodynamic limit, by an intermediate step where the quenched free 
energy was shown to be subadditive in the size of the system.

Moreover, still by interpolation on families of Gaussian random variables, 
the first mentioned result was extended to give a rigorous proof that the 
expression given by the Parisi \textit{Ansatz} is also a lower bound for 
the quenched free energy of the system, uniformly in the size \cite{Grepli}. 
The method gives not only the bound, but also the explicit form of the 
correction in a quite involved form. As a recent and very important result, 
along the task of facing the challenge, Michel Talagrand has been able to 
dominate these correction terms, showing that they vanish in the 
thermodynamic limit. This milestone achievement was firstly announced in a 
short note \cite{Tb}, containing only a synthetic sketch of the proof, and then 
presented with all details in a long paper to be published on Annals of 
Mathematics \cite{Topus}.

The interpolation method is also at the basis of the far reaching 
generalized variational principle proven by Michel Aizenman, Robert Sims and 
Shannon Starr in \cite{ASS}.

In our presentation, we will try to be as self-contained as possible. We 
will give all definitions, explain the basic structure of the interpolation 
method, and show how some of the results are obtained. We will concentrate 
mostly on questions connected with the free energy, its properties of 
subadditivity, the existence of the infinite volume limit, and the replica 
bounds.

For the sake of comparison, and in order to provide a kind of warm up, we 
will recall also some features of the standard elementary mean field model 
of ferromagnetism, the so called Curie-Weiss model. We will concentrate 
also here on the free energy, and systematically exploit elementary 
comparison and interpolation arguments. This will show the strict analogy 
between the treatment of the ferromagnetic model and the developments in 
the mean field spin glass case. Basic roles will be played in the two 
cases, but with different expressions, by positivity and convexity 
properties.

The organization of the paper is as follows. In Section 2, we introduce 
the ferromagnetic model and discuss behavior and properties of the free 
energy in the thermodynamic limit, by emphasing, in this very elementary 
case, the comparison and interpolation methods that will be also 
exploited, in a different context, in the spin glass case. 

Section 3 is devoted to the basic features of the mean field spin glass 
models, by introducing all necessary definitions.

In Section 4, we introduce the Gaussian comparison and interpolation 
method, by giving simple applications to the existence of the infinite 
volume limit of the quenched free energy \cite{GTthermo}, and to the proof of general 
variational bounds, by following the useful strategy developed in \cite{ASS}.

Section 5 will briefly recall the main features of the Parisi 
representation, and will state the main theorem concerning the free energy.

Finally, Section 6 will be devoted to conclusions and outlook for future 
developments.

It is a pleasure to thank the Organizing Committee, and in particular 
Professor Ari Laptev, for the kind invitation to talk in a such stimulating 
cultural athmosphere.

\section {The mean field ferromagnetic model. Structure and results.}

The mean field ferromagnetic model is among the simplest models of 
statistical mechanics. However, it contains very interesting features, in 
particular a phase transition, characterized by spontaneous magnetization, 
at low temperatures. 
We refer to standard textbooks, for example \cite{Stanley}, for a full 
treatment, and a complete appreciation of the model in the frame of the 
theory of ferromagnetism. Here we consider only some properties of the free 
energy, easily obtained through comparison methods.

The generic configuration of the  mean field ferromagnetic model is defined 
through Ising spin variables
$\sigma_{i}=\pm 1$,  attached to each site $i=1,2,\dots,N$.

The Hamiltonian of the model, in some external field of strength $h$,  
is given by the mean field expression
\begin{equation}\label{H}
H_N(\sigma,h)=-{1\over N}\sum_{(i,j)}\sigma_i\sigma_j
-h\sum_{i}\sigma_i.
\end{equation}
Here, the first sum extends to all $N(N-1)/2$ site couples, 
an the second to all sites.

For a given inverse temperature $\beta$, let us now introduce the 
partition function $Z_{N}(\beta,h)$ and 
the free energy per site
$f_{N}(\beta,h)$,  
according to the well known definitions
\begin{eqnarray}\label{Z}
&&Z_N(\beta,h)=\sum_{\sigma_1\dots\sigma_N}\exp(-\beta H_N(\sigma,h)),\\
\label{f}
&&-\beta f_N(\beta,h)=N^{-1} E\log Z_N(\beta,h).
\end{eqnarray}

It is also convenient to define the average spin magnetization
\begin{equation}\label{m}
m={1\over N}\sum_{i}\sigma_i.
\end{equation}

Then, it is immediately seen that the Hamiltonian in (\ref{H}) can be 
equivalently written as
\begin{equation}\label{Hm}
H_N(\sigma,h)=-{1\over2} N m^2
-h\sum_{i}\sigma_i,
\end{equation}
where an unessential constant term has been neglected. In fact we have
\begin{equation}\label{HHm}
\sum_{(i,j)}\sigma_i\sigma_j = {1\over2} \sum_{i,j;i\ne j}\sigma_i\sigma_j=
{1\over2} N^2 m^2 - {1\over2} N, 
\end{equation}
where the sum over all couples has been equivalently written as one half the sum 
over all $i,j$ with $i\ne j$, and the diagonal terms with $i=j$ have been added 
and 
subtracted out. Notice that they give a constant because $\sigma_i^2=1$.

Therefore, the partition function in (\ref{Z}) can be equivalently substituted 
by the expression
\begin{equation}\label{Z'}
Z_N(\beta,h)=\sum_{\sigma_1\dots\sigma_N}\exp(- {1\over2} \beta N m^2)
\exp(\beta h \sum_{i}\sigma_i), 
\end{equation} 
which will be our starting point.

Our interest will be in the $\lim_{N\to\infty} N^{-1} \log Z_N(\beta,h)$.
To this purpose, let us establish the important subadditivity property, 
holding for the splitting of the big N site system in two smaller $N_1$ 
site and $N_2$ site systems, respectively, with $N=N_1+N_2$,
\begin{equation}\label{sub}
\log Z_N(\beta,h)\le \log Z_{N_1}(\beta,h) + \log Z_{N_2}(\beta,h).
\end{equation}
The proof is very simple. Let us denote, in the most natural way, by 
$\sigma_1,\dots,\sigma_{N_1}$ the spin variables for the first subsystem, 
and by $\sigma_{N_1 +1},\dots,\sigma_{N}$ the $N_2$ spin variables of the 
second subsystem. Introduce also the subsystem magnetizations $m_1$ and 
$m_2$, by adapting the definition (\ref{m}) to the smaller systems, in such 
a way that
\begin{equation}\label{m'}
N m= N_1 m_1 + N_2 m_2.
\end{equation}
Therefore, we see that the large system magnetization $m$ is the linear 
convex combination of the smaller system ones, according to the obvious
\begin{equation}\label{m''}
m= {N_1\over N} m_1 + {N_2\over N} m_2.
\end{equation}
Since the mapping $m\to m^2$ is convex, we have also the general bound, 
holding for all values of the $\sigma$ variables
\begin{equation}\label{m2}
m^2 \le {N_1\over N} m_1^2 + {N_2\over N} m_2^2.
\end{equation}
Then, it is enough to substitute the inequality in the definition 
(\ref{Z'}) of 
$Z_N(\beta,h)$, and recognize that we achieve factorization with 
respect to the two subsystems, and therefore the inequality 
$Z_N\le Z_{N_1} Z_{N_2}$. So we have established (\ref{sub}). From 
subadditivity, the existence of the limit follows by a simple argument, as 
explained for example in \cite{ruelle}. In fact, we have
\begin{equation}\label{lim}
\lim_{N\to\infty}N^{-1}\log Z_N(\beta,h)= \inf_{N}N^{-1}\log Z_N(\beta,h).
\end{equation}
Now we will calculate explicitely this limit, by introducing an order 
parameter $M$, a trial function, and an appropriate variational scheme.
In order to get a lower bound, we start from the elementary inequality 
$m^2 \ge 2 m M - M^2$, holding for any value of $m$ and $M$. By inserting 
the inequality in the definition (\ref{Z'}) we arrive at a factorization of 
the sum over $\sigma$'s. The sum can be explicitely calculated, and we 
arrive immediately to the lower bound, uniform in the size of the system,
\begin{equation}\label{lb}
N^{-1}\log Z_N(\beta,h) \ge \log2 + \log\cosh\beta(h+M) - {1\over2}\beta 
M^2,
\end{equation}
holding for any value of the trial order parameter $M$. Clearly it is 
convenient to take the supremum over $M$. Then we establish the optimal 
uniform lower bound
\begin{equation}\label{lb'}
N^{-1}\log Z_N(\beta,h) \ge \sup_M(\log2 + \log\cosh\beta(h+M) - {1\over2}\beta 
M^2).
\end{equation}

It is simple to realize that the supremum coincides with the limit as 
$N\to\infty$. To this purpose we follow the following simple 
procedure. Let us consider all possible values of the variable $m$. There 
are $N+1$ of them, corresponding to any number $K$ of possible spin flips, 
starting from a given $\sigma$ configuration, $K=0,1,\dots,N$. Let us 
consider the trivial decomposition of the identity, holding for any $m$,
\begin{equation}\label{sum}  
1=\sum_M \delta_{m M},
\end{equation}
where $M$ in the sum runs over the $N+1$ possible values of $m$, and 
$\delta$ is Kroneker delta, beeing equal to $1$ if $M=N$, and zero 
otherwise. Let us now insert (\ref{sum}) in the definition (\ref{Z'}) of the 
partition function inside the sum over $\sigma$'s, and invert the two sums. 
Because of the forcing 
$m=M$ given by the $\delta$, we can write $m^2 = 2 m M - M^2$ inside the 
sum. Then if we neglect the $\delta$, by using the trivial $\delta \le 1$, 
we have un upper bound, where the sum over $\sigma$'s can be explicitily 
performed as before. Then it is enough to take the upper bound with 
respect to $M$, and consider that there are $N+1$ terms in the now trivial 
sum over $M$, in order to arrive at the upper bound
\begin{equation}\label{ub}
N^{-1}\log Z_N(\beta,h) \le \sup_M(\log2 + \log\cosh\beta(h+M) - {1\over2}\beta 
M^2) + N^{-1}\log (N+1).
\end{equation}
Therefore, by going to the limit as $N\to\infty$, we can collect all our 
results in the form of the following theorem giving the full 
characterization of the thermodynamic limit of the free energy.
\begin{theorem}
\label{tlim}
For the mean field ferromagnetic model we have
\begin{eqnarray}    
&&\lim_{N\to\infty}N^{-1}\log Z_N(\beta,h) =  \inf_N N^{-1}\log Z_N(\beta,h)\\
&&=\sup_M(\log2 + \log\cosh\beta(h+M) - {1\over2}\beta M^2).
\end{eqnarray}
\end{theorem}
This ends our discussion about the free energy in the ferromagnetic model. 
Now we are ready to attack the much more difficult spin glass model. But 
it will be surprising to see that, by following a simple extension of the 
methods here described, we will arrive to similar results.

\section{The basic definitions for the mean field spin glass model}

As in the ferromagnetic case, the generic configuration of the mean field spin 
glass model is defined 
through Ising spin variables
$\sigma_{i}=\pm 1$,  attached to each site $i=1,2,\dots,N$. 

But now there is an external quenched disorder 
given by the $N(N-1)/2$ independent and identical distributed random
variables $J_{ij}$, defined for each couple of sites. For the sake of simplicity,
we assume each $J_{ij}$ to be a centered
unit Gaussian with averages $E(J_{ij})=0$, $ E(J_{ij}^2)=1$. By quenched 
disorder we mean that the $J$ have a kind of stochastic external influence 
on the system, without partecipating to the thermal equilibrium.   

Now the Hamiltonian of the model, in some external field of strength $h$,  
is given by the mean field expression
\begin{equation}\label{Hsg}
H_N(\sigma,h,J)=-{1\over\sqrt{N}}\sum_{(i,j)}J_{ij}\sigma_i\sigma_j
-h\sum_{i}\sigma_i.
\end{equation}
Here, the first sum extends to all site couples, an the second to all sites. 
Notice the $\sqrt{N}$ necessary to ensure a good thermodynamic behavior to 
the free energy.

For a given inverse temperature $\beta$, let us now introduce the 
disorder dependent
partition function $Z_{N}(\beta,h,J)$ and
the quenched average of the free energy per site
$f_{N}(\beta,h)$,   
according to the  definitions
\begin{eqnarray}\label{Zsg}
&&Z_N(\beta,h,J)=\sum_{\sigma_1\dots\sigma_N}\exp(-\beta H_N(\sigma,h,J)),\\
\label{fsg}
&&-\beta f_N(\beta,h)=N^{-1} E\log Z_N(\beta,h,J). 
\end{eqnarray}
Notice that in (\ref{fsg}) the average $E$ with respect to the external noise is 
made \textit{after} the $\log$ is taken. This procedure is called quenched 
averaging. It represents the physical idea that the external noise does not 
partecipate to the thermal equilibrium. Only the $\sigma$'s are thermalized.

For the sake of simplicity, it is also convenient to write the partition
function in the following equivalent form. First of all let us introduce 
a family of centered Gaussian random variables $\kappa(\sigma)$, indexed by the  
configurations $\sigma$, and characterized by the covariances
\begin{equation}\label{cov}
E\bigl(\kappa(\sigma)\kappa(\sigma^\prime)\bigr)=q^2(\sigma,\sigma^\prime),
\end{equation}
where $q(\sigma,\sigma^\prime)$ are the overlaps between two generic 
configurations, defined by
\begin{equation}\label{overlap}
q(\sigma,\sigma^\prime)=
N^{-1}\sum_{i}\sigma_i\sigma^{\prime}_i,
\end{equation}
with the obvious bounds
$-1\le q(\sigma,\sigma^\prime)\le 1$, and the normalization 
$q(\sigma,\sigma) = 1$.
Then, starting from the definition (\ref{Hsg}), it is immediately seen that 
the partition function in (\ref{Zsg}) can be also written, by neglecting 
unessential constant terms, in the form  
\begin{equation}\label{Zsg'}     
Z_N(\beta,h,J)=
\sum_{\sigma_1\dots\sigma_N}\exp(\beta \sqrt{N\over2}\kappa(\sigma))
\exp(\beta h \sum_{i}\sigma_i),
\end{equation}    
which will be the starting point of our treatment.

\section{Gaussian comparison and applications}

Our basic comparison argument will be based on the following very simple 
theorem.
\begin{theorem}
\label{comparison}
Let $U_i$ and $\hat U_i$, for $i=1,\dots,K$, be independent families of 
centered Gaussian random variables, whose covariances satisfy the 
inequalities for generic configurations
\begin{equation}\label{dominance}
E(U_i U_j) \equiv S_{ij} \ge E(\hat U_i \hat U_j) \equiv \hat S_{ij},
\end{equation}
and the equalities along the diagonal
\begin{equation}\label{diagonal}
E(U_i U_i) \equiv S_{ii} = E(\hat U_i \hat U_i) \equiv \hat S_{ii},
\end{equation}
then for the quenched averages we have the inequality in the opposite sense
\begin{equation}\label{basic}
E\log \sum_i w_i\exp(U_i) \le E\log \sum_i w_i\exp(\hat U_i),
\end{equation} 
where the 
$w_i\ge 0$ are the same in the two expressions. 
\end{theorem}
The proof is extremely simple and amounts to a straigthforward 
calculation. In fact, let us consider the interpolating expression
\begin{equation}\label{int}
E\log \sum_i w_i\exp(\sqrt{t}U_i + \sqrt{1-t}\hat U_i), 
\end{equation}
where $0\le t \le 1$. Clearly the two expressions under comparison 
correspond to the values $t=0$ and $t=1$ respectively. By taking the 
derivative with respect to $t$, and then integrating by parts with respect 
to the Gaussian variables, we immediately see that the interpolating 
function is nonincreasing in $t$, and the theorem follows. On the other 
hand, considerations of this kind are present in the mathematical 
literature of some years ago. Two typical references are \cite{Joag} 
and \cite{kahane}.

We give here some striking applications of the basic comparison Theorem.
In \cite{GTthermo} we have given a very simple proof of a long waited 
result, about the convergence of the free energy per site in the 
thermodynamic limit. Let us show the argument. Let us consider a system of 
size $N$ and two smaller systems of sizes $N_1$ and $N_2$ respectively, 
with $N=N_1+N_2$, as before in the ferromagnetic case. Let us now compare
\begin{equation}\label{ElogZ}     
E\log Z_N(\beta,h,J)=E\log
\sum_{\sigma_1\dots\sigma_N}\exp(\beta \sqrt{N\over2}\kappa(\sigma))
\exp(\beta h \sum_{i}\sigma_i),
\end{equation}
with
\begin{eqnarray}
\nonumber
&&E\log
\sum_{\sigma_1\dots\sigma_N}\exp(\beta 
\sqrt{N_1\over2}\kappa^{(1)}(\sigma^{(1)})
\exp(\beta 
\sqrt{N_2\over2}\kappa^{(2)}(\sigma^{(2)})
\exp(\beta h \sum_{i}\sigma_i) \\
\label{Z1Z2}
&&\equiv E\log Z_{N_1}(\beta,h,J)
+E\log Z_{N_2}(\beta,h,J) ,
\end{eqnarray}
where $\sigma^{(1)}$ are the $(\sigma_i,\ i=1,\dots,N_1)$, and
$\sigma^{(2)}$ are the $(\sigma_i,\ i=N_1+1,\dots,N)$. Covariances for
$\kappa^{(1)}$ and $\kappa^{(2)}$ are expressed as in (\ref{cov}), but now the 
overlaps are substituted with the partial overlaps of the first and second 
block, $q_1$ and $q_2$ respectively. It is very simple to apply the comparison 
theorem. All one has to do is to observe that the obvious
\begin{equation}\label{q}
N q= N_1 q_1 + N_2 q_2,
\end{equation}
analogous to (\ref{m'}), implies, as in (\ref{m2}),
\begin{equation}\label{q2}
q^2 \le {N_1\over N} q_1^2 + {N_2\over N} q_2^2.
\end{equation}
Therefore, the comparison gives the superaddivity property, to be compared 
with (\ref{sub}),
\begin{equation}\label{super}
E\log Z_N(\beta,h,J)\ge E\log Z_{N_1}(\beta,h,J) + E\log Z_{N_2}(\beta,h,J).
\end{equation}
From the superaddivity property the existence of the limit follows in the 
form
\begin{equation}\label{limsg}
\lim_{N\to\infty}N^{-1}E\log Z_N(\beta,h,J)= \sup_{N}N^{-1}E\log Z_N(\beta,h,J),
\end{equation}
to be compared with (\ref{lim}).

The second application is in the form of the Aizenman-Sims-Starr 
generalized variational principle. Here, we will need to introduce some 
auxiliary system. The denumerable configuration space is given by the values of 
$\alpha=1,2,\dots$. We introduce also a probability measure $w_{\alpha}$ 
for the $\alpha$ system, and suitably defined overlaps between two generic configurations 
$p(\alpha,\alpha^{\prime})$, with $p(\alpha,\alpha)=1$. A family of 
centered Gaussian random variables $\hat\kappa(\alpha)$, now indexed by the  
configurations $\alpha$, will be defined by the covariances
\begin{equation}\label{covhat}
E\bigl(\hat\kappa(\alpha) \hat\kappa(\alpha^\prime)\bigr)=
p^2(\alpha,\alpha^\prime).
\end{equation}  
We will need also a family of centered Gaussian random variables 
$\eta_{i}(\alpha)$, indexed by the sites $i$ of our original system and the 
configurations $\alpha$ of the auxiliary system, so that
\begin{equation}\label{eta}
E\bigl(\eta_i(\alpha) \eta_{i^\prime}(\alpha^\prime)\bigr)=
\delta_{i i^\prime} p(\alpha,\alpha^\prime).
\end{equation}

Both the probability measure $w_\alpha$, and the overlaps 
$p(\alpha,\alpha^{\prime})$ could depend on some additional external 
quenched noise, that does not appear explicitely in our notation.

In the following, we will denote by $E$ averages with respect to all 
random variables involved.

In order to start the comparison argument, we will consider firstly the 
case where the two  $\sigma$ and $\alpha$ systems are not coupled, so to 
appear factorized in the form
\begin{eqnarray}
\nonumber
&&E\log
\sum_{\sigma_1\dots\sigma_N}\sum_{\alpha} w_{\alpha}
\exp(\beta \sqrt{N\over2}\kappa(\sigma))
\exp(\beta \sqrt{N\over2}\hat\kappa(\alpha))
\exp(\beta h \sum_{i}\sigma_i) \\
\label{first}
&&\equiv E\log Z_N(\beta,h,J) +
E\log
\sum_{\alpha} w_{\alpha}
\exp(\beta \sqrt{N\over2}\hat\kappa(\alpha)).
\end{eqnarray}

In the second case the $\kappa$ fields are suppressed and the coupling 
between the two systems will be taken in a very simple form, by allowing 
the $\eta$ field to act as an external field on the $\sigma$ system. In 
this way the $\sigma$'s appear as factorized, and the sums can be 
explicitely performed. The chosen form for the second term in the 
comparison is
\begin{eqnarray}
\nonumber
&&E\log
\sum_{\sigma_1\dots\sigma_N}\sum_{\alpha} w_{\alpha}
\exp(\beta  \sum_{i}\eta_i(\alpha)\sigma_i)
\exp(\beta h \sum_{i}\sigma_i)\\
\label{second}
&& \equiv N\log2 +
E\log \sum_{\alpha} w_{\alpha}(c_1 c_2 \dots c_N), 
\end{eqnarray}
where we have defined
\begin{equation}\label{ci}
c_i = \cosh \beta(h+\eta_i(\alpha)),
\end{equation}
as arising from the sums over $\sigma$'s.

Now we apply the comparison Theorem. In the first case, the covariances 
involve the sums of squares of overlaps
\begin{equation}\label{squares}
{1\over2} \bigl(q^2(\sigma,\sigma^\prime)+p^2(\alpha,\alpha^\prime)\bigr). 
\end{equation}
In the second case, a very simple calculation shows that the covariances 
involve the overlap products 
\begin{equation}\label{product}
q(\sigma,\sigma^\prime) p(\alpha,\alpha^\prime). 
\end{equation}
Therefore, the comparison is very easy and, by collecting all expressions, 
we end up with the useful estimate, as in \cite{ASS}, holding for any 
auxiliary system as defined before,
\begin{eqnarray}
\label{variational} 
&&N^{-1} E\log Z_N(\beta,h,J) \le\\
\nonumber
&&\log2 + N^{-1} \bigl(E\log \sum_{\alpha} w_{\alpha}(c_1 c_2 \dots c_N)-
E\log
\sum_{\alpha} w_{\alpha}
\exp(\beta \sqrt{N\over2}\hat\kappa(\alpha))\bigr).
\end{eqnarray}

\section{The Parisi representation for the free energy}

We refer to the original paper \cite{P}, and to the extensive 
review given in \cite{MPV}, for the general motivations, and the 
derivation of the broken replica  {\it Ansatz}, in the frame of the 
ingenious replica trick. Here we limit ourselves to a synthetic 
description of its general structure, independently from the replica trick

First of all, let us introduce the convex space ${\cal X}$ of the functional 
order parameters $x$, as nondecreasing functions of the auxiliary variable 
$q$,
both $x$ and $q$ taking
values on the interval $[0,1]$, {\it i.e.}
\begin{equation}
\label{x}
{\cal X}\ni x : [0,1]\ni q \rightarrow x(q) \in [0,1].
\end{equation}
Notice that we call $x$ the function, and $x(q)$ its values.
We introduce a metric on ${\cal X}$ through the $L^{1}([0,1], dq)$ norm, where 
$dq$ is the Lebesgue measure.

For our purposes, we will consider the case of piecewise constant functional order 
parameters, characterized by an integer $K$, and two sequences $q_0, q_1, 
\dots, q_K$, $m_1, m_2, \dots, m_K$ of numbers satisfying
\begin{equation}
\label{qm}
0=q_0\le q_1 \le \dots \le q_{K-1} \le q_K=1,\,\,\, 0\le m_1 \le m_2 \le \dots 
\le m_K \le 1,
\end{equation}
such that
\begin{eqnarray}
\nonumber
x(q)=m_1 \,\,\mbox{for}\,\, 0=q_0\le q < q_1,\,\,\, 
x(q)=m_2 &\mbox{for}& q_1\le q < q_2,\\
\label{xpcws}
\ldots,
 x(q)=m_K &\mbox{for}& q_{K-1}\le q \le q_K.
\end{eqnarray}
In the following, we will find convenient to define also $m_0\equiv 0$, 
and $m_{K+1}\equiv 1$. The replica symmetric case of Sherrington and 
Kirkpatrick corresponds to 
\begin{equation}
\label{replicas}
K=2,\,\, 
q_1=\bar q, \,\, m_1=0,\,\, m_2=1.
\end{equation}

Let us now introduce the function $f$, with values $f(q,y;x,\beta)$, of 
the variables $q\in[0,1]$, $y\in R$, depending also on the functional order 
parameter $x$, and on the inverse temperature $\beta$, defined as the 
solution of the nonlinear antiparabolic equation 
\begin{equation}
\label{antipara}
(\partial_q f)(q,y)+
{1\over2}(\partial_y^2 f)(q,y)+{1\over2}x(q)({\partial_y f})^2(q,y)=0,
\end{equation}
with final condition
\begin{equation}
\label{final}
f(1,y)=\log\cosh(\beta y).
\end{equation}
Here, we have stressed only the dependence of $f$ on $q$ and $y$.

It is very simple to integrate Eq.~(\ref{antipara}) when $x$ is piecewise 
constant. In fact, consider $x(q)=m_a$, for $q_{a-1}\le q \le q_{a}$, 
firstly with $m_a>0$. Then, 
it is immediately seen that the correct solution of Eq.~(\ref{antipara}) in 
this interval, with the right final boundary condition at $q = q_{a}$, is 
given by
\begin{equation}
\label{solution}
f(q,y)=\frac{1}{m_a}\log \int\exp\bigl({m_{a} f(q_a,y+z\sqrt{q_a-q})}\bigr)\,d\mu(z),
\end{equation}
where $d\mu(z)$ is the centered unit Gaussian measure on the real line. On 
the other hand, if $m_a=0$, then (\ref{antipara}) loses the nonlinear part 
and the solution is given by
\begin{equation}
\label{solution0}
f(q,y)= \int f(q_a,y+z\sqrt{q_a-q})\,d\mu(z),
\end{equation}
which can be seen also as deriving from (\ref{solution}) in the limit $m_a \to 0$.
Starting from the last interval $K$, and using (\ref{solution}) iteratively on 
each interval, we easily get the solution of (\ref{antipara}), 
(\ref{final}), in the case of piecewise order parameter $x$, as in
(\ref{xpcws}).

Now we introduce the following important definitions.
The trial auxiliary function, associated to a given mean field spin glass 
system, as described in Section 3, depending on the functional order 
parameter $x$, is defined as
\begin{equation}
\label{trial}
\log 2 + 
f(0,h;x,\beta)-\frac{\beta^2}{2}\int_{0}^{1} 
q\, x(q)\,dq.
\end{equation}
Notice that in this expression the function $f$ appears evaluated at $q=0$, 
and $y=h$, where $h$ is the value of the external magnetic field. This 
trial expression shoul be considered as the analog of that appearing in 
(\ref{lb}) for the ferromagnetic case.

The Parisi spontaneously broken replica symmetry expression for the free 
energy is given by the definition
\begin{equation}\label{parisi}
-\beta f_P(\beta,h) \equiv \inf_x \bigl(\log 2 + 
f(0,h;x,\beta)-\frac{\beta^2}{2}\int_{0}^{1} 
q\, x(q)\,dq \bigr),
\end{equation}
where the infimum is taken with respect to all functional order parameters 
$x$. 
Notice that the infimum appears here, as compared to the supremum in the 
ferromagnetic case.

In \cite{Grepli}, by exploiting a kind of generalized comparison argument, 
involving a suitably defined interpolation function, we have established 
the following important result.

\begin{theorem}
\label{main}
For all values of the inverse temperature $\beta$, and the external 
magnetic field $h$, and for any functional order parameter $x$, the 
following bound holds
$$
N^{-1}E\log Z_{N}(\beta,h,J)\le \log 2 + 
f(0,h;x,\beta)-\frac{\beta^2}{2}\int_{0}^{1} 
q\, x(q)\,dq ,
$$
uniformly in $N$. Consequently, we have also
$$
N^{-1}E\log Z_{N}(\beta,h,J)\le\inf_x \bigl(\log 2 + 
f(0,h;x,\beta)-\frac{\beta^2}{2}\int_{0}^{1} 
q\, x(q)\,dq \bigr),
$$
uniformly in $N$.
\end{theorem}

However, this result can be understood also in the frame of the 
generalized variational principle established by Aizenman-Sims-Starr and 
described before.

In fact, one can easily show that there exist an $\alpha$ systems such that
$$
N^{-1}E\log \sum_{\alpha} w_{\alpha}(c_1 c_2 \dots c_N)\equiv f(0,h;x,\beta),
$$
$$
N^{-1}E\log
\sum_{\alpha} w_{\alpha}
\exp(\beta \sqrt{N\over2}\hat\kappa(\alpha))\equiv\frac{\beta^2}{2}\int_{0}^{1} 
q\, x(q)\,dq ,
$$
uniformly in $N$. This result stems from previous work of Derrida, Ruelle, 
Neveu,  
Bolthausen, Sznitman, Aizenman, Talagrand, Bovier, and others, and in a sense is 
implicit in the treatment given in \cite{MPV}. We plan to deal with this 
important representation in a forthcoming note.

We see that the estimate in Theorem~\ref{main} are also a consequence of 
the generalized variational principle.

Up to this point we have seen how to obtain upper bounds. The problem 
arises whether, as in the ferromagnetic case, we can also get lower 
bounds, so to shrink the thermodynamic limit to the value given by the 
$\inf_x$ in Theorem~\ref{main}. After a short announcement in \cite{Tb}, 
Michel Talagrand wrote an extended paper \cite{Topus}, 
to appear on Annals of Mathematics, where the complete proof of the control 
of the lower bound is firmly established. We refer to the original paper 
for the complete details of this remarkable achievement. About the 
methods, here we only recall that in \cite{Grepli} we have given also the 
corrections to the bounds appearing in Theorem~\ref{main}, albeit in a 
quite complicated form. Talagrand, with 
great courage, has been able to establish that these corrections do in 
fact vanish in the thermodynamic limit.

In conclusion, we can establish the following extension of Theorem~\ref{tlim} 
to spin glasses.
\begin{theorem}
\label{tlimsg}
For the mean field spin glass model we have
\begin{eqnarray}
&&\lim_{N\to\infty}N^{-1}E\log Z_N(\beta,h,J) =  \sup_N N^{-1}E\log 
Z_N(\beta,h,J)\\
&&=\inf_x \bigl(\log 2 + 
f(0,h;x,\beta)-\frac{\beta^2}{2}\int_{0}^{1} 
q\, x(q)\,dq \bigr).
\end{eqnarray}
\end{theorem}

\section{Conclusion and outlook for future developments}

As we have seen, in these last few years there has been an impressive progress in the 
understanding of 
the mathematical structure of spin glass models, mainly due to the 
systematic exploration of comparison and interpolation methods. 
However many important problems are still open. The most important one is 
to establish rigorously the full hierarchical ultrametric organization of 
the overlap distributions, as appears in Parisi theory, 
and to fully understand the decomposition in pure states of the glassy phase, at 
low temperatures.

Moreover, is would be important to extend these methods to other important 
disordered models as for example neural networks. Here the difficulty is 
that the positivity arguments, so essential in comparison methods, do not 
seem to emerge naturally inside the structure of the theory.

We plan to report on these problems in future works. 

\vspace{.5cm}
{\bf Acknowledgments}

We gratefully acknowledge useful conversations with Michael Aizenman, 
Pierluigi Contucci,
Giorgio Parisi and Michel Talagrand. The strategy explained in this paper 
grew out from a 
systematic exploration of comparison and interpolation methods, developed in 
collaboration with Fabio Lucio Toninelli.

This work was supported in part by MIUR 
(Italian Minister of Instruction, University and Research), 
and by INFN (Italian National Institute for Nuclear Physics).

\end{document}